\documentclass{svproc}

\usepackage{booktabs}
\usepackage{url}

\usepackage{graphicx}
\usepackage{float}

\begin{document}
\mainmatter             
\title{Efficient Audio-Visual Event Recognition via Knowledge Distillation and Dynamic INT8 Quantization of a Hybrid Cross-Attention Network}
\titlerunning{Efficient Audio-Visual Event Recognition}  
%
\author{Parinaz Binandeh Dehaghani\inst{1} \and Danilo Pena\inst{2} \and A. Pedro Aguiar\inst{3}}

\authorrunning{}

\tocauthor{Parinaz Binandeh Dehaghani, Danilo Pena, and A. Pedro Aguiar}

\institute{
University of Porto, Porto, Portugal\\
\email{up202100618@edu.fe.up.pt}
\and
ResoSight, Montreal, Canada\\
\email{danilo.pena@resosight.com}
\and
University of Porto, Porto, Portugal\\
\email{pedro.aguiar@fe.up.pt}
}




\maketitle

\begin{abstract}
Audio-visual event recognition (AVER) has achieved significant performance improvements through transformer-based multimodal architectures. However, the high computational complexity, large memory footprint, and inference cost of these models hinder their deployment on edge and resource-constrained devices. This paper presents an efficient compression framework for hybrid cross-attention-based audio-visual event recognition by combining architectural model compression, knowledge distillation, and dynamic INT8 quantization. A high-capacity teacher model integrates VideoMAE for visual representation learning, the Audio Spectrogram Transformer (AST) for audio feature extraction, and a hybrid cross-attention fusion network for multimodal feature integration. A lightweight student model is constructed by reducing the hidden feature dimension, the number of attention heads, and the feed-forward network size while preserving the overall network architecture. The student model is trained using knowledge distillation to effectively transfer discriminative knowledge from the teacher. Finally, dynamic INT8 post-training quantization is applied to further reduce the model size for efficient deployment. Experimental results on the Audio-Visual Event (AVE) dataset show that the proposed framework reduces the number of trainable parameters in the multimodal fusion module by \textbf{59.06\%}, with only a \textbf{2.14\%} decrease in classification accuracy compared with the teacher model. Furthermore, dynamic INT8 quantization reduces the model size from \textbf{10.71 MB} to \textbf{2.04 MB} while maintaining competitive recognition performance. These results demonstrate that the proposed framework provides an effective trade-off between recognition accuracy and computational efficiency, making it a promising solution for deployment on resource-constrained edge AI platforms.

\keywords{Audio-Visual Event Recognition, Knowledge Distillation, Model Compression, Dynamic INT8 Quantization, Cross-Attention, VideoMAE, Audio Spectrogram Transformer, Edge AI}
\end{abstract}

\section{Introduction}
AVER aims to recognize events by jointly analyzing visual scenes and their corresponding acoustic signals. Unlike unimodal approaches, multimodal learning exploits the complementary information contained in both audio and video streams, leading to more robust event understanding in complex real-world environments. Owing to its ability to capture richer semantic information, AVER has attracted considerable attention in recent years and has found applications in intelligent surveillance, autonomous driving, robotics, smart cities, multimedia retrieval, and human--computer interaction \cite{tian2018audio,owens2016ambient}.
Recent advances in transformer architectures have significantly improved multimodal representation learning. In particular, the AST has demonstrated remarkable performance for audio understanding by modeling spectrogram patches with self-attention mechanisms \cite{gong2021ast}, while VideoMAE has achieved state-of-the-art visual representation learning through self-supervised masked video modeling \cite{tong2022videomae}. Combined with cross-modal attention mechanisms, these powerful feature extractors enable modern AVER systems to effectively model semantic correlations between audio and visual streams, substantially outperforming conventional convolutional neural network (CNN)-based approaches.

Despite these advances, the superior recognition performance of transformer-based AVER models comes at the cost of high computational complexity, large memory footprints, and expensive attention operations. In particular, hybrid cross-attention fusion networks introduce additional computational overhead because they explicitly model bidirectional interactions between audio and visual representations. Consequently, deploying these models on resource-constrained platforms, such as edge AI devices, autonomous robots, wearable systems, and intelligent transportation platforms, remains a significant challenge. While numerous studies have focused on designing increasingly sophisticated multimodal fusion architectures, comparatively little attention has been devoted to developing efficient compression strategies specifically tailored to hybrid cross-attention-based audio-visual transformer networks.
Model compression has emerged as an effective approach for reducing computational complexity while preserving recognition performance. Existing techniques include network pruning, quantization, low-rank approximation, and knowledge distillation \cite{han2015deep,hinton2015distilling,jacob2018quantization}. However, most existing compression methods are designed for generic vision or language models and are directly applied as post-processing techniques without considering the architectural characteristics of multimodal fusion networks. Consequently, they do not explicitly preserve the multimodal interaction mechanisms that are essential for hybrid cross-attention-based audio-visual event recognition.
To address this limitation, this paper proposes an \emph{architecture-aware compression framework} specifically designed for hybrid cross-attention-based audio-visual event recognition. Rather than compressing the entire multimodal pipeline, the proposed framework targets the trainable multimodal fusion network while preserving the representational capability of the pretrained VideoMAE and AST backbones. The pretrained VideoMAE and AST backbone encoders are used only for offline feature extraction and remain unchanged. Consequently, the proposed compression framework focuses on the trainable multimodal fusion and classification network, which represents the deployable reasoning component of the overall AVER system. The proposed teacher model integrates VideoMAE for visual representation learning, AST for audio feature extraction, and a Stable Hybrid Cross-Attention fusion network for multimodal reasoning. A lightweight student network is then constructed by systematically reducing the hidden feature dimension, the number of attention heads, and the feed-forward network capacity while maintaining the original multimodal fusion topology. Knowledge distillation is subsequently employed to transfer the multimodal reasoning capability of the teacher to the compressed student, followed by dynamic INT8 post-training quantization to further reduce the deployment memory footprint without additional retraining.
Experimental results on the AVE benchmark demonstrate that the proposed framework reduces the number of trainable parameters by 59.06\% while incurring only a 2.14\% decrease in classification accuracy. Furthermore, dynamic INT8 quantization reduces the student model size from 10.71~MB to 2.04~MB, demonstrating that the proposed architecture-aware compression framework achieves an effective balance between recognition accuracy and deployment efficiency for edge AI applications.
The main contributions of this paper are summarized as follows:

\begin{itemize}

\item We propose an architecture-aware compression framework specifically designed for hybrid cross-attention-based audio-visual event recognition that preserves the multimodal reasoning capability of the original fusion architecture while substantially reducing the computational complexity and model size of the trainable multimodal fusion and classification network.

\item We develop a lightweight student architecture that preserves the original multimodal fusion topology while substantially reducing the hidden feature dimension, attention heads, and feed-forward network capacity.

\item We integrate response-based knowledge distillation and dynamic INT8 post-training quantization into a unified compression pipeline to effectively transfer multimodal knowledge and further reduce storage requirements without additional retraining.

\item Extensive experiments on the AVE benchmark demonstrate that the proposed framework reduces the number of trainable parameters by 59.06\% and the quantized model size by more than 80\%, while maintaining competitive recognition performance, making it suitable for deployment on resource-constrained edge AI platforms.

\end{itemize}

\section{Related Work}

\subsection{Audio-Visual Event Recognition}

AVER has attracted increasing attention because many real-world events are characterized by complementary visual and acoustic information. Early studies mainly relied on CNNs to independently extract audio and visual features, followed by late-fusion or feature-concatenation strategies for event classification. The AVE dataset introduced by Tian \emph{et al.}~\cite{tian2018audio} established the first large-scale benchmark for audio-visual event localization and recognition, stimulating extensive research in multimodal learning. Subsequent works demonstrated that jointly modeling audio and visual modalities significantly improves robustness compared with unimodal systems, particularly when one modality is noisy, ambiguous, or partially unavailable~\cite{arandjelovic2018objects,owens2016ambient,xu2020cross,baltruvsaitis2018multimodal}.
To better exploit cross-modal relationships, attention-based fusion mechanisms have gradually replaced conventional feature concatenation. Cross-attention enables one modality to selectively attend to informative representations from the other modality, allowing more effective multimodal interaction and feature alignment. Recent transformer-based architectures have further advanced this direction by modeling long-range temporal dependencies and semantic correlations across modalities~\cite{gong2021ast,tong2022videomae,dosovitskiy2021vit,vaswani2017attention}. Several studies have shown that hybrid attention mechanisms combining modality-wise and temporal attention achieve superior recognition performance compared with traditional fusion approaches, making transformer-based cross-attention the dominant paradigm for modern audio-visual event recognition~\cite{wang2024cmbt,geng2023cmpt,xu2020cross}.

\subsection{Transformer-Based Audio-Visual Representation Learning}

Transformer architectures have substantially advanced both audio and visual representation learning. For audio analysis, the AST introduced by Gong \emph{et al.}~\cite{gong2021ast} demonstrated that spectrogram patches can be effectively modeled using self-attention without convolutional layers, achieving state-of-the-art performance on several audio classification benchmarks. In the visual domain, VideoMAE~\cite{tong2022videomae} employs masked video modeling to learn powerful spatio-temporal representations through self-supervised learning, providing robust visual embeddings for downstream recognition tasks.
Recent audio-visual recognition frameworks increasingly combine transformer-based feature extractors with cross-modal fusion modules. By jointly learning interactions between visual and acoustic representations, these hybrid architectures outperform conventional CNN-based fusion methods. Nevertheless, the improved recognition accuracy is accompanied by increased computational complexity, higher memory consumption, and larger model sizes, limiting their deployment on edge and embedded platforms.

\subsection{Model Compression for Deep Neural Networks}

Model compression has become an active research area for improving the deployment efficiency of deep neural networks. Among the most widely adopted techniques are network pruning, quantization, and knowledge distillation. Pruning removes redundant parameters from a trained network to reduce computational complexity, while quantization represents model parameters using lower numerical precision to decrease memory usage and accelerate inference~\cite{han2015deep,jacob2018quantization}. Knowledge distillation, introduced by Hinton \emph{et al.}~\cite{hinton2015distilling}, transfers knowledge from a high-capacity teacher model to a compact student network, enabling lightweight models to preserve much of the teacher's predictive capability.

Several recent studies have explored model compression for audio understanding and edge AI applications, demonstrating that knowledge distillation and quantization can substantially reduce model complexity while maintaining competitive recognition performance. However, most existing works focus on unimodal audio or visual models, or investigate individual compression techniques in isolation. Recent surveys also highlight the growing interest in combining multiple compression strategies to improve deployment efficiency while preserving model accuracy~\cite{gou2021knowledge,moslemi2024survey}. Similar observations are reported in recent work on environmental sound classification, where coordinated compression pipelines are shown to be more effective than applying individual techniques independently.
Despite these advances, efficient compression of hybrid transformer-based audio-visual architectures remains relatively underexplored. Existing AVER studies primarily emphasize improving recognition accuracy through increasingly sophisticated multimodal fusion strategies, whereas comparatively less attention has been devoted to reducing model complexity for deployment on resource-constrained devices. Motivated by this gap, this paper proposes a lightweight teacher-student framework that combines architectural compression, knowledge distillation, and dynamic INT8 quantization for hybrid cross-attention-based audio-visual event recognition.

\section{Methodology}
\begin{figure}[t]
    \centering
    \includegraphics[
        width=\linewidth,
        trim=0.6cm 0.4cm 0.6cm 0.4cm,
        clip
    ]{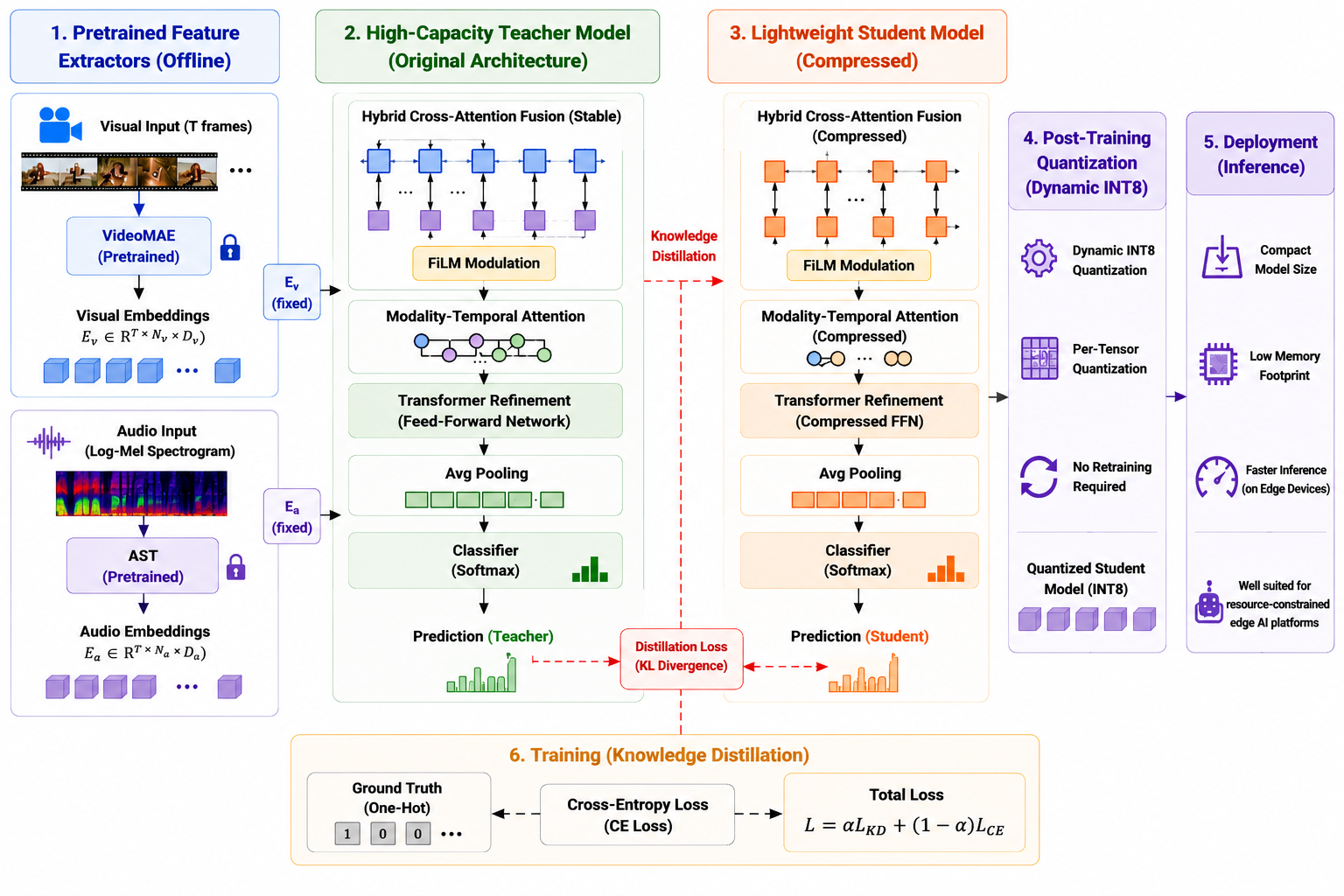}
    \caption{Overview of the proposed teacher--student compression framework for hybrid cross-attention-based audio-visual event recognition.}
    \label{fig:framework}
\end{figure}

During preprocessing, visual and audio embeddings are extracted once using pretrained VideoMAE and AST encoders and stored as cached feature representations. During training and inference, the proposed framework operates exclusively on these cached embeddings. Therefore, all reported parameter counts, model sizes, and compression ratios correspond to the trainable multimodal fusion and classification network rather than the fixed backbone feature extractors.
This section presents the proposed compression framework for efficient audio-visual event recognition. The framework consists of four major stages: (1) training a high-capacity teacher model, (2) constructing a lightweight student architecture, (3) transferring knowledge from the teacher to the student using knowledge distillation, and (4) applying dynamic INT8 post-training quantization for deployment efficiency. Fig.~\ref{fig:framework} illustrates the overall workflow of the proposed framework.


\subsection{Overall Framework}

The proposed framework aims to preserve the recognition capability of a transformer-based hybrid audio-visual model while substantially reducing its computational complexity. First, a large teacher model is trained using supervised learning on the AVE dataset. The teacher consists of VideoMAE for visual representation learning, the AST for audio feature extraction, and a hybrid cross-attention fusion module for multimodal interaction.
A lightweight student model is then derived by reducing the complexity of the transformer fusion network while maintaining the same overall architecture. During training, the student simultaneously learns from the ground-truth labels and the soft predictions generated by the teacher network through knowledge distillation. Finally, dynamic INT8 post-training quantization is applied to the trained student model to further decrease memory consumption without requiring additional retraining.

\subsection{Teacher Architecture}

The teacher network adopts a hybrid transformer-based multimodal architecture. Video features are extracted using the pretrained VideoMAE encoder, which produces temporal visual representations from sampled video segments. Audio features are obtained using the pretrained AST, which converts log-Mel spectrograms into high-level acoustic embeddings.
To effectively integrate both modalities, the extracted features are first projected into a common embedding space through learnable linear projection layers. The projected representations are then refined using a lightweight bidirectional cross-attention module. Specifically, visual features attend to audio features while audio features simultaneously attend to visual representations, allowing complementary information to be exchanged between modalities.
Unlike conventional transformer fusion architectures, the proposed cross-attention module employs residual scaling initialized near zero. This initialization preserves the stability of the original feature representations during early training while allowing the model to gradually learn cross-modal interactions. Finally, the refined representations are aggregated using the Modality Temporal Attention module, producing a compact multimodal feature vector for event classification.

\subsection{Compressed Student Network}

The student model preserves the overall network topology of the teacher while significantly reducing its computational complexity. Instead of modifying the backbone feature extractors, compression is performed within the transformer fusion module.

Three architectural modifications are introduced:

\begin{itemize}
\item The hidden embedding dimension is reduced from 512 to 256.
\item The number of multi-head attention heads is reduced from eight to four.
\item The feed-forward network dimension is reduced from 1024 to 512.
\end{itemize}
These modifications considerably reduce the number of trainable parameters while preserving the overall multimodal fusion strategy. Since both teacher and student share the same network topology, knowledge transfer becomes more effective during distillation.

\subsection{Knowledge Distillation}

Knowledge distillation is employed to transfer the discriminative capability of the teacher model to the compressed student network. During training, the teacher parameters remain frozen, while only the student network is optimized.
The student is trained using a weighted combination of the conventional cross-entropy loss and a distillation loss computed from the teacher's soft predictions. The teacher logits are softened using a temperature parameter $T$, enabling the student to capture inter-class relationships that are not available from hard labels alone.
The total loss is defined as

\begin{equation}
\mathcal{L}
=
(1-\alpha)\mathcal{L}_{CE}
+
\alpha T^2
\mathcal{L}_{KD},
\end{equation}
where $\alpha$ controls the contribution of the distillation loss and $T$ denotes the temperature parameter.

\subsection{Dynamic INT8 Quantization}

After the student model is fully trained, dynamic INT8 post-training quantization is applied to improve deployment efficiency. Unlike quantization-aware training, this approach does not require retraining the network.
Only the linear layers of the transformer are quantized from 32-bit floating-point precision to 8-bit integer representations. During inference, weights are stored in INT8 format while activations remain dynamically quantized according to their runtime distribution.
This strategy substantially reduces the model storage requirements and memory bandwidth while preserving most of the recognition accuracy. Since transformer architectures contain numerous fully connected layers, dynamic quantization provides a particularly effective compression technique for this application.

\subsection{Training Objective}

The teacher network is first optimized using supervised cross-entropy learning. After convergence, its parameters are frozen and used to guide the student model through knowledge distillation.
The compressed student is trained until convergence using the combined loss function described above. Finally, the best-performing checkpoint is selected according to the validation accuracy and quantized using dynamic INT8 post-training quantization. The resulting compressed model is evaluated on the AVE test set in terms of classification accuracy, weighted precision, weighted recall, weighted F1-score, model size, and number of trainable parameters.

\section{Experimental Setup}

Experiments were conducted on the AVE dataset introduced by Tian \emph{et al.}~\cite{tian2018audio}. The AVE dataset comprises 4,143 unconstrained video clips collected from YouTube, covering 28 event categories, including musical instruments, animals, vehicles, and human activities. Each video is approximately 10 seconds long and contains synchronized visual and audio streams. Following the official evaluation protocol, the dataset was divided into predefined training, validation, and testing subsets.
During preprocessing, each video was uniformly divided into ten one-second temporal segments to preserve the temporal evolution of audio-visual events. Visual and audio features were extracted independently using the pretrained VideoMAE and AST encoders, respectively. To improve training efficiency, the extracted features were computed offline and cached, eliminating repeated backbone inference during model training while maintaining synchronized audio-visual representations for each temporal segment.
The teacher model employs the proposed Stable Hybrid Cross-Attention architecture, which integrates bidirectional cross-attention with the original MAFnet modality-temporal attention mechanism for multimodal feature fusion. The pretrained VideoMAE and AST backbones remained frozen throughout training, and only the fusion and classification modules were optimized. To obtain a lightweight student model, the hidden feature dimension was reduced from 512 to 256, the number of attention heads from 8 to 4, and the feed-forward network dimension from 1024 to 512, while preserving the overall network architecture and a single transformer refinement layer.
Knowledge transfer from the teacher to the student was performed using response-based knowledge distillation. The optimization objective combines the standard cross-entropy loss with the Kullback--Leibler divergence between the softened output distributions of the teacher and student networks. A temperature of $T=4$ and a weighting coefficient of $\alpha=0.8$ were adopted to balance the hard-label and soft-label supervision. After distillation, the compressed student model was further optimized using post-training dynamic INT8 quantization, where all linear layers were quantized to 8-bit integer precision while the remaining operations continued to use floating-point arithmetic.

All models were implemented in PyTorch and optimized using the AdamW optimizer with an initial learning rate of $1\times10^{-4}$ and a weight decay of $1\times10^{-4}$. Training was performed for a maximum of 80 epochs using a batch size of four and label smoothing of 0.05. A ReduceLROnPlateau learning-rate scheduler was employed to adaptively decrease the learning rate when the validation performance plateaued. Furthermore, an enhanced early stopping strategy jointly monitored validation accuracy and validation loss to mitigate overfitting while preserving the best-performing model.

\section{Results and Discussion}

This section evaluates the effectiveness of the proposed compression framework on the AVE dataset. The experiments compare three models: the original teacher network, the compressed student network trained via knowledge distillation, and the dynamically quantized INT8 student model. The comparison considers both recognition performance and deployment efficiency.

\begin{figure}
    \centering
    \includegraphics[width=1\linewidth]{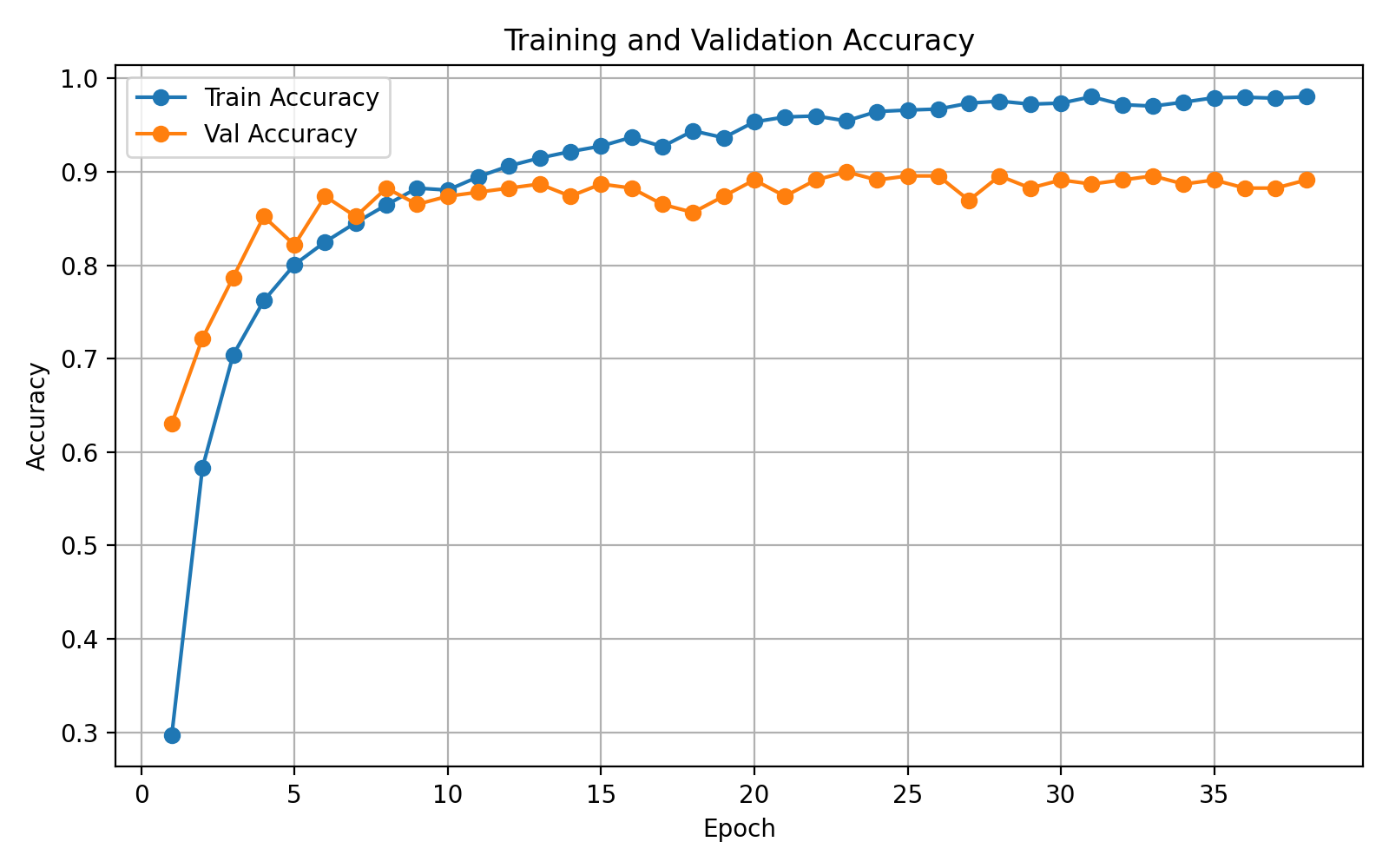}
    \caption{Training and validation accuracy curves of the proposed compressed student model on the AVE dataset.}
    \label{fig:accuracy}
\end{figure}

\begin{figure}
    \centering
    \includegraphics[width=1\linewidth]{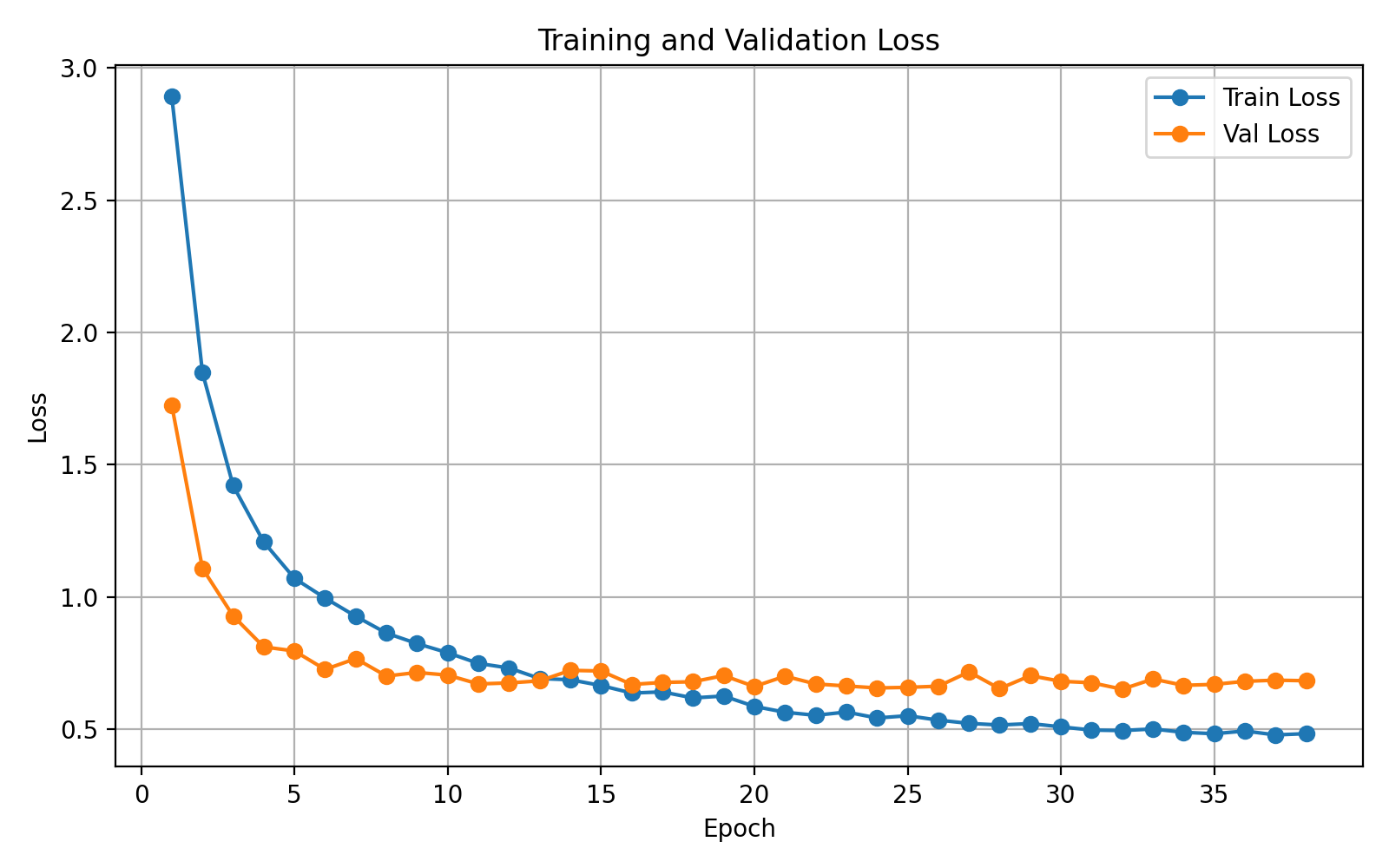}
    \caption{Training and validation loss curves of the proposed student model during knowledge distillation.}
    \label{fig:loss}
\end{figure}

\subsection{Overall Performance Comparison}

\begin{figure}[t]
    \centering
    \includegraphics[
        width=\linewidth,
        trim=2cm 1.5cm 2cm 1cm,
        clip
    ]{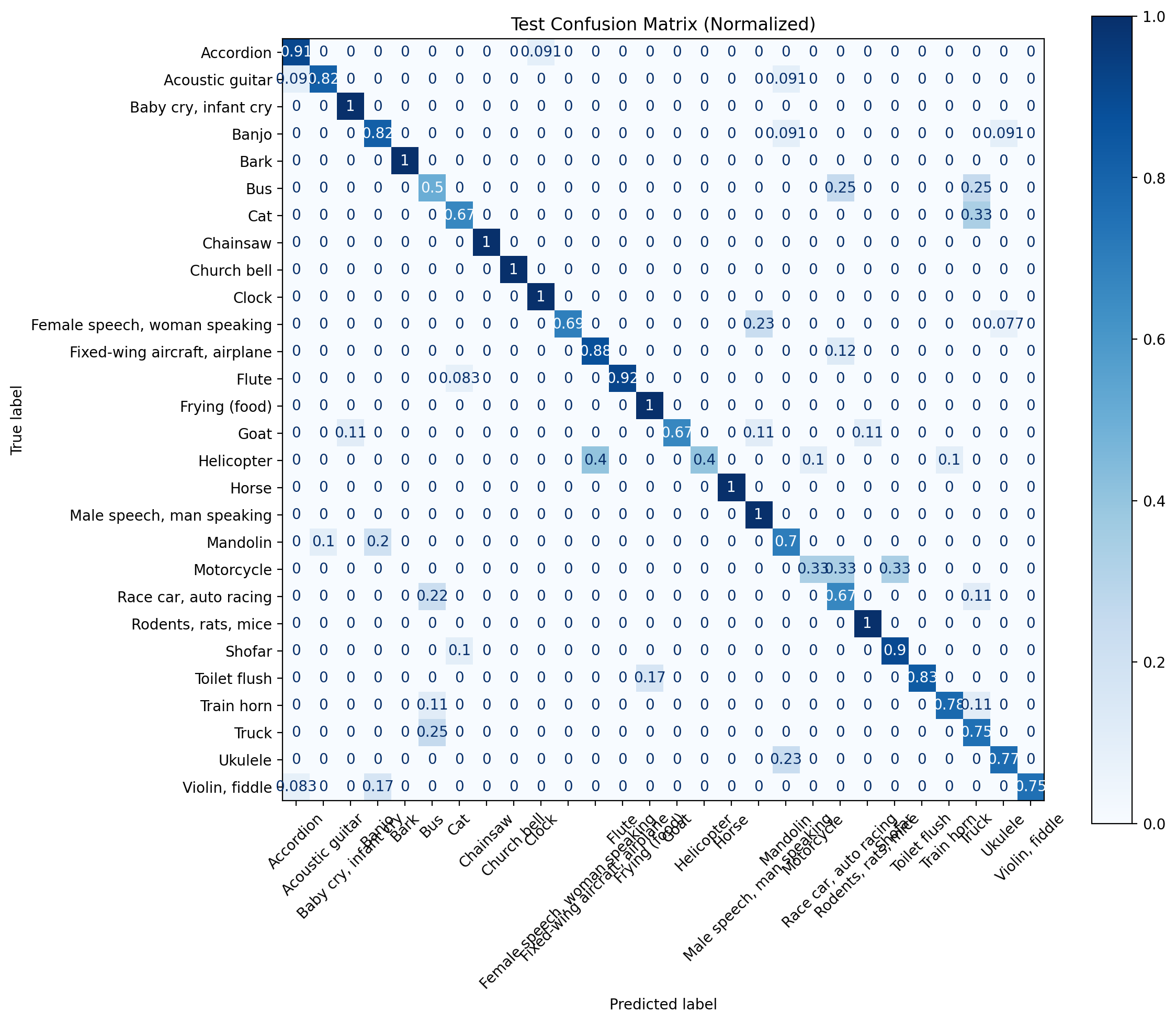}
    \caption{Normalized confusion matrix of the proposed compressed student model on the AVE test set.}
    \label{fig:student}
\end{figure}

Table~\ref{tab:comparison} summarizes the experimental results for the three evaluated models. The teacher model achieves the highest recognition performance, obtaining a test accuracy of 84.19\% with a weighted F1-score of 84.17\%. After architectural compression and knowledge distillation, the student model achieves an accuracy of 82.05\%, corresponding to an absolute performance decrease of only 2.14\%. Despite this small reduction in accuracy, the student network reduces the number of trainable parameters from 6.86 million to 2.81 million, representing a 59.06\% reduction.
The effectiveness of the proposed compression framework is further illustrated by the training curves shown in Fig.~\ref{fig:accuracy} and Fig.~\ref{fig:loss}. As shown in Fig.~\ref{fig:accuracy}, both the training and validation accuracies increase steadily throughout the optimization process, indicating stable convergence of the compressed student model during knowledge distillation. Meanwhile, Fig.~\ref{fig:loss} demonstrates that both the training and validation losses decrease consistently and stabilize after approximately 20 epochs, suggesting that the student network successfully learns the discriminative knowledge transferred from the teacher without exhibiting significant overfitting.
These results indicate that the proposed architecture-aware compression framework effectively transfers multimodal knowledge from the teacher model to the compressed student network. The relatively small accuracy degradation demonstrates that most of the multimodal representation capability is preserved despite the substantially smaller architecture, confirming that the proposed student design provides an effective trade-off between recognition performance and model complexity. 
Fig.~\ref{fig:student} presents the normalized confusion matrix of the compressed student model on the AVE test set. Most event categories are correctly classified, as indicated by the strong diagonal structure of the matrix. Only a limited number of confusions occur between visually or acoustically similar events, while the majority of classes achieve high recognition rates. These results further confirm that the proposed architecture-aware compression framework effectively preserves the multimodal discriminative capability of the teacher network despite reducing the number of trainable parameters by 59.06\%.

\subsection{Compression Analysis}

Besides improving computational efficiency, the proposed framework significantly reduces the storage requirements of the model. The teacher network occupies 26.16~MB of memory, whereas the student model requires only 10.71~MB. Similarly, the checkpoint size decreases from 78.53~MB to 32.19~MB.
The overall compression ratio between the teacher and student models reaches 2.44$\times$, while the number of trainable parameters is reduced by nearly 60\%. These results demonstrate that reducing the hidden embedding dimension, the number of attention heads, and the feed-forward network dimension provides an effective architectural compression strategy without introducing substantial performance degradation.

\subsection{Effect of Dynamic INT8 Quantization}

To further improve deployment efficiency, dynamic INT8 post-training quantization was applied to the compressed student model. After quantization, the model size decreased from 10.71~MB to only 2.04~MB, while the checkpoint size was reduced from 32.19~MB to 4.25~MB.
Although quantization introduced a slight decrease in classification accuracy, the INT8 model still achieved a competitive accuracy of 81.20\%, corresponding to an overall accuracy reduction of only 2.99\% relative to the original teacher model. These results demonstrate that dynamic quantization provides substantial memory savings while preserving most of the recognition capability learned through knowledge distillation.
Interestingly, the measured inference latency of the dynamically quantized model was slightly higher than that of the floating-point student model. This behavior is primarily attributed to the relatively small size of the compressed network and the overhead associated with dynamic quantization during CPU execution. Nevertheless, the considerable reduction in storage requirements makes the quantized model highly suitable for deployment on memory-constrained edge devices.

\subsection{Discussion}

The experimental results demonstrate that the proposed framework achieves an effective balance between recognition accuracy and computational efficiency. Unlike conventional compression approaches that rely solely on pruning or quantization, the proposed method combines architectural compression, knowledge distillation, and dynamic post-training quantization into a unified pipeline.
The architectural simplifications considerably reduce the computational complexity of the multimodal fusion network, while knowledge distillation compensates for the loss of representational capacity by transferring semantic information from the teacher model. Finally, dynamic INT8 quantization further compresses the model for practical deployment without requiring additional training.
Overall, the proposed framework reduces the number of trainable parameters by 59.06\%, decreases the model size by more than 80\% after quantization, and preserves over 96\% of the teacher model's classification accuracy. These results demonstrate that the proposed framework is a promising solution for deployment in resource-constrained edge AI applications, offering an effective balance between recognition performance and model efficiency. 


\begin{table*}[t]
\centering
\caption{Performance comparison of the teacher model, compressed student model, and dynamically quantized INT8 student model on the AVE dataset.}
\label{tab:comparison}
\renewcommand{\arraystretch}{1.2}
\begin{tabular}{lccc}
\toprule
\textbf{Metric} & \textbf{Teacher} & \textbf{Student (KD)} & \textbf{Student + Dynamic INT8} \\
\midrule
Test Accuracy (\%)                     & \textbf{84.19} & 82.05 & 81.20 \\
Test Loss                              & \textbf{0.889} & 0.936 & 0.963 \\
Precision (\%)                         & \textbf{86.57} & 84.96 & 84.41 \\
Recall (\%)                            & \textbf{84.19} & 82.05 & 81.20 \\
Weighted F1-score (\%)                 & \textbf{84.17} & 82.13 & 81.06 \\
\midrule
Parameters                             & 6,855,711 & 2,806,815 & 2,806,815 \\
Trainable Parameters                   & 6,855,711 & 2,806,815 & 2,806,815 \\
Model Size (MB)                        & 26.16 & 10.71 & \textbf{2.04} \\
Checkpoint Size (MB)                   & 78.53 & 32.19 & \textbf{4.25} \\
\midrule
Compression Ratio (Teacher $\rightarrow$ Student) & -- & $2.44\times$ & -- \\
Parameter Reduction (\%)               & -- & 59.06 & -- \\
Accuracy Drop from Teacher (\%)        & -- & 2.14 & 2.99 \\
Checkpoint Size Reduction vs. Student (\%) & -- & -- & 86.78 \\
\bottomrule
\end{tabular}
\end{table*}


\section{Conclusion}

This paper presents an efficient compression framework for hybrid cross-attention-based audio-visual event recognition by integrating architectural model compression, knowledge distillation, and dynamic INT8 post-training quantization. The proposed framework preserves the multimodal fusion capability of a high-capacity teacher model while significantly reducing the computational complexity of the student network. Specifically, the student architecture was constructed by reducing the hidden embedding dimension, the number of attention heads, and the feed-forward network size, while knowledge distillation was employed to transfer discriminative multimodal knowledge from the teacher model. Dynamic INT8 quantization was subsequently applied to further reduce the storage requirements of the compressed model without additional retraining.
Experimental results on the AVE dataset demonstrated that the proposed framework effectively balances recognition performance and computational efficiency. Compared with the original teacher model, the compressed student model achieved a 59.06\% reduction in trainable parameters while incurring only a 2.14\% decrease in classification accuracy. Furthermore, dynamic INT8 quantization reduced the model size from 10.71~MB to 2.04~MB, providing substantial memory savings while maintaining competitive recognition performance. The proposed framework significantly reduces the memory footprint of the trainable multimodal reasoning component while preserving the representational capability of the pretrained backbone encoders, making it well suited for deployment in resource-constrained edge AI applications.
Future work will investigate more advanced compression techniques, including structured pruning, quantization-aware training, and adaptive knowledge distillation strategies. In addition, the proposed framework will be evaluated on larger-scale audio-visual benchmarks and real-world edge computing platforms to further validate its effectiveness in practical deployment scenarios.

\bibliographystyle{plain}
\bibliography{bibtex/ref}
\end{document}